\documentclass[intlimits,twoside,a4paper]{article}

\usepackage{amsmath,amssymb}
\usepackage{graphicx}
\usepackage{wrapfig}

\usepackage[T2A]{fontenc}
\usepackage[cp1251]{inputenc}

\usepackage[eqsecnum]{cmpj2}

\issue{2013}{16}{1}{13703}

\doinumber{10.5488/CMP.16.13703}

\title[Dipole glass parameter behaviour]%
{Dipole glass parameter behaviour for ferro-antiferroelectric solid mixtures}
\author[M.A. Korynevskii, V.B. Solovyan]{M.A. Korynevskii\refaddr{label1,label2,label3},
        V.B. Solovyan\refaddr{label1}}
\addresses{
\addr{label1} Institute for Condensed Matter Physics of the National Academy of Sciences of Ukraine, \\
1 Svientsitskii St., 79011 Lviv, Ukraine
\addr{label2} Lviv Polytechnic National University, 12 Bandera St., 79013 Lviv, Ukraine
\addr{label3} Institute of Physics of the University of Szczecin, 15~Wielkopolska St., 70451 Szczecin, Poland
}

\date{Received  May 25, 2012, in final form September 24, 2012}
\authorcopyright{M.A. Korynevskii, V.B. Solovyan, 2013}

\newcommand{\non}{\nonumber \\}

\newcommand{\be}{\begin{equation}}
\newcommand{\ee}{\end{equation}}
\newcommand{\bea}{\begin{eqnarray}}
\newcommand{\eea}{\end{eqnarray}}

\newcommand{\lp}{\left (}
\newcommand{\rp}{\right )}
\newcommand{\lb}{\left [}
\newcommand{\rb}{\right ]}
\newcommand{\lbr}{\left \{}
\newcommand{\rbr}{\right \}}
\newcommand{\ld}{\left .}

\begin{document}

\maketitle

\begin{abstract}
A new definition of a dipole glass parameter for ferro-antiferroelectric solid mixtures is suggested. It is constructed on the nearest
neighbours pair correlation functions for
interacting dipole momenta. The behaviour of the dipole glass parameter  is calculated and discussed.
\keywords ferroelectrics, antiferroelectrics, mixed systems, correlation functions, dipole glass
\pacs 76.30.K, 77.80.Bh
\end{abstract}

\section{Introduction}

The problem of a dipole glass state in frustrated ferroelectrics is difficult both in experimental detection and for theoretical
interpretation. It still remains  actual, because this state of non-ordering system at low temperatures is widely observed in different materials,
for example in dielectrics with
non-central ions, relaxors, mixed ferro-antiferroelectric compounds~\cite{1,2,3,4}.
Dipole glass state is a characteristic one for structurally complex systems and corresponds to non-equilibrium thermodynamic phases in them.
The most important property of the system and a simple criterion for detection of the process of dipole glass transition is a dispersion of dielectric
characteristics. Therefore, in experimental investigations, the dynamical methods for the study of real and imaginary parts
of dielectric susceptibility behaviour under high and low frequencies are used. Naturally, this state does not depend on the external field effect and must be observed in
a static case.

On the other hand, due to a competition between different types of interparticle interactions in a complex crystalline matrix there is a possibility of different local states coexisting with their non-equilibrium occupation. Those states correspond to a local minima of the free
energy of the system with high potential
barriers between them. In the thermodynamic limit ($N\rightarrow \infty$, $V\rightarrow \infty$, $N/V=\textrm{const}$), the relaxation time for transition from some state to another one tends to
infinity. As far as the system is close to such a local minimum, it remains there for a long time. These systems are not ergodic and their
theoretical description is quite complicated. The dispersion of dielectric characteristics of the systems in a dipole glass state is caused by
the stimulated transitions from a certain energy state to another one. So, the dynamic behaviour of the investigated system can be illustrated by their microscopic structure,
or in other words, by their static correlation functions.

The experimental study of dipole glass phases in frustrated ferroelectrics is presented mostly \linebreak in papers devoted to mixed ferro-antiferroelectric
compounds based on the KDP-type crystals \linebreak{} [K$_n$(NH$_4$)$_{1-n}$H$_2$PO$_4$\,, \, Rb$_n$(NH$_4$)$_{1-n}$H$_2$PO$_4$\,, \, Rb$_n$(NH$_4$)$_{1-n}$D$_2$PO$_4$\,, \, Rb$_n$(NH$_4$)$_{1-n}$H$_2$AsO$_4$].
The most fundamental of them, in our opinion, are dielectric investigations~\cite{5,6,7,8,9,10,11}. In all of them, the temperature behaviour of real ($\epsilon'$)
and imaginary ($\epsilon''$) parts of dielectric permittivity at low and high frequencies looks similar for concentrations $0.2<n<0.8$. Namely,
the existence of smooth ($\epsilon'$) or sharp ($\epsilon''$) peaks at temperatures about 50~K indicate some ordering processes in the system.
Since no long-range ordered phases in this range of concentrations were observed, only an abnormal increase of short-range correlations is possible there.
This state was called a dipole glass phase.

The EPR~\cite{12,13}, NMR~\cite{14} and X-ray~\cite{15} structural investigations have shown the existence of significant local fields in ferro-antiferroelectric mixtures
at low temperatures. Those fields are connected with strong short-range correlations between charged particles (protons).

\looseness=-1 The theoretical study of mixed ferro-antiferroelectric systems is based on the use of a replica method for calculation of the basic thermodynamic functions.
This method was first suggested in the theory of diluted magnetic or mixed ferro-antiferromagnetic systems~\cite{16,17,18,19,20}. As for mixed hydrogen bonded
ferro-antiferroelectric compounds, the Ising model with a random transverse and longitudinal fields was explored. The Gaussian distribution for random interactions and the
corresponding fields was used in~\cite{21,22}. A cluster theory of mixed KDP-type ferroelectrics, including the effect of piezoelectric coupling, was proposed in~\cite{23,24,25}.
The phase diagrams, the behaviour of thermodynamic functions, long-range and dipole glass order parameters were obtained and discussed. Most of those results
are in good agreement with the experiment.

In our previous papers~\cite{26,27,28}, a new approach to the study of hydrogen bonded ferro-antiferro\-electric solid mixtures of Rb$_n$(NH$_4$)$_{1-n}$H$_2$AsO$_4$ (RADA)
type was suggested. It is based on the concept of a decisive role of microscopic structure of a system (single and pair correlation functions)
in the formation of a dipole glass phase. The cluster mean field approximation with replica method was used for their calculation. So, the ``concentration-phase transition temperature'' phase diagram  was obtained in good agreement with the experimental one. The areas of the existence of dipole glass phases were interpreted as multiplication of pair correlation
functions at a certain concentration for one of the components at relatively low temperature. Since the pair correlation functions characterize
a short-range order in the system, the dipole glass state is determined by different possibilities of short-range arrangement of interacting particles,
i.e., by their stable and metastable states all over the sample.

As for a strict definition of a dipole glass parameter, there appear to be some problems. Usually~\cite{16} it is defined as a mean value of the dipole momentum square.
However, in this case, the multi-stage character of dipole glass phase and the real interparticle correlations are not taken into account.
Moreover, the temperature behaviour of dipole glass parameter does not react adequately upon passing the ``freezing'' point (the point
of the dipole glass phase appearance). This parameter remains non-equal to zero even in the paraelectric phase where no glass state can be observed.
A more appropriate definition of this parameter and the static interpretation of a dipole glass phase will be presented in the present paper.

\section{Model. Analytical results}

It is well known that~\cite{29} the ferroelectric RbH$_2$AsO$_4$ and antiferroelectric NH$_4$H$_2$AsO$_4$ single crystals are isomorphic with tetragonal lattice
$D_{2d}^{12}$ space group symmetry in paraelectric phase. Below 110~K in the first crystal and below 213~K in the second one, a ferroelectric and antiferroelectric ordering are
correspondingly  observed. The axis of the orientation of dipole moments in RbH$_2$AsO$_4$ is directed along $0Z$, but in NH$_4$H$_2$AsO$_4$ it is
directed along $0X$ $(0Y)$.

Solid mixtures of both crystals due to frustration of ferro- and antiferroelectric types of interactions demonstrate a possibility of a new ``ordered''
dipole glass phase appearance. In this phase, no type of long-range ordering is preferential and only short-range correlations between particles remain.
Dipole glass phase is realized for intermediate concentration ($0.50<n<0.80$) and at relatively low temperatures (below 100~K). The physical reason for the existence of these phases are numerous arrangements of nearest neighbours (particles) in a disordered ferro-antiferroelectric system.

We will regard a regular KDP-type crystalline lattice, each site of which is randomly occupied by rubidium atoms or ammonium groups,
that is by Rb or NH$_4$. The effective Hamiltonian of this mixed system is taken in the form~\cite{28,30}:
\bea
H &=& - \frac{1}{2} \sum_{i,j} \sum_{mm'} \lb V_{ij}^{mm'}\hat n_i \hat n_j
S_{im}^z S_{jm'}^z + U_{ij}^{mm'}(1-\hat n_i) (1-\hat n_j)
S_{im}S_{jm'}^x\rb \non
&&+ \frac{1}{2} \sum_i \sum_m \lb E\cos \Theta \hat n_i S^z_{im} + E \sin \Theta
(1-\hat n_i) S_{im}^x \rb,
\label{eq:2.1}
\eea
where both $V_{ij}^{mm'}$ and $U_{ij}^{mm'}$ are the intensities of
interaction between $m$-th dipole particle in $i$-th site with
$m'$-th dipole particle in $j$-th site when dipole moments are oriented
along $z$-axis and along $x$-axis, correspondingly. $S^z$ and $S^x$
are projections of the unit classical vector $\vec S$, the site
occupation operators $\hat n_i$ have the following eigenvalues:
\be
\label{eq:2.2}
n_i =
\left\{ \begin{array}{ll}
1, \qquad \text{when $i$-th site is occupied by Rb atom},\\
0, \qquad \text{when $i$-th site is occupied by NH$_4$ group}.
\end{array}\right.
\ee
Here, $E$ is the external electric field, $\theta$ is the angle between $z$-axis and the direction of this field.

To take into account the random distribution of dipoles oriented along
$z$-axis (Rb-AsO$_4$ groups) and dipoles oriented along $x$-axis (NH$_4$-AsO$_4$ groups) we have used the replica method~\cite{18} for the configuration averaging
of thermodynamic functions. Since there are two types of dipoles having ferroelectric and antiferroelectric orientation,  we must correspondingly introduce two
types of replica variables
$\sigma_{im}^k$ for $S_{im}^z$ and
$\xi_{im}^k$ for $S_{im}^x$. The following property of a binomial
random variable $\chi_c$ will be used. If the probability of $\chi_c$ is $\mathrm{Pr}\lbr
\chi_c=r\rbr=\lp \begin{smallmatrix}c \\ r \end{smallmatrix} \rp p^r
q^{c-r}$, then the expectation value of $\exp \lbr a\chi_c \rbr$ is
equal to $\lp p \re^a +q \rp^c$, which behaves like $1+c \ln \lp p \re^a +q
\rp$, when the analytic continuation for $c\rightarrow 0$ is performed.
In the context of the investigated system, the $p$ and $q$ are as follows: $p=n$,
$q=1-n$, $n$ is a concentration (density) of
Rb atoms.

Since the unit cell for KDP-type crystal contains two formula
units (there are also two sublattices below the phase transition
point~\cite{29}), the numbers $m$, $m'$ run from 1 to 2.
So, we consider the next form for the interactions between all particles in the crystalline lattice: the strict
two-particle interaction for particles belonging to
the same crystalline lattice site and the self-consistent mean field approximation for
particles from different sites, the so-called two-particle cluster approximation~\cite{31}.

With the accuracy up to the second order of symmetric replica
expansion and using two-particle cluster approximation, the
following expression for the free energy of the system investigated has
been obtained:
\bea
\label{eq:2.3}
F &=& -  \frac{1}{\beta} \ln \lbr 4 \lb \re^{-A_1}+\re^{A_1} \cosh \lp 2 B_1 + \beta E \cos \Theta \rp\rb\right. \non
&&\times\ld \lb \re^{A_2}\cosh \lp \beta E \sin\Theta\rp + \re^{-A_2}\cosh (2B_2)\rb\rbr + \frac{1}{\beta} C\,.
\eea
Here,
\bea
\label{eq:2.4}
 A_1 &=& J_1+J_2
\left\{\tilde V_f^2-\tilde V_{af}^2 + \left[ 1 + \left(
\tilde V_f-\tilde V_{af}\right)^2\right] g_z -
\tilde U g_x\right\},\non
A_2 &=& \left(\tanh \beta V - J_1\right)
\tilde U + J_2\left\{ \tilde U^2_f- \tilde U^2_{af}+ \left[
\tilde U^2 + \left( \tilde U_f - \tilde U_{af}\right)^2 \right]
g_x - \tilde U g_z \right\},\non
B_1 &=& \lb J_1 \tilde V_f + 4J_2 \left(
\tilde V_f^2 P^2 -3 \tilde V_f
\tilde U_{af} q^2\right)\rb P\,,\non
B_2 &=& \lb \left( \tanh \beta V - J_1\right)
\tilde U_{af} + 4J_2 \left( \tilde U_{af}^2 q^2 - 3\tilde V_f \tilde U_{af} P^2
\right) \rb q\,,\eea

\bea
C &=& J_1 \tilde V_f P^2
+ \left( \tanh \beta V - J_1\right)\hat U_{af} q^2 +
\frac{J_2}{2} \left[ -1 -\tilde U + \left( g_z -
\tilde U g_x \right)^2 \right.\non
&&
+ \left( \tilde V_f + \tilde V_{af} \right)^2 + \left(
\tilde V_f - \tilde V_{af}\right)^2 g_z^2 + \left(\tilde U_f+\tilde U_{af} \right)^2
+ \ld \left( \tilde U_f + \tilde U_{af} \right)^2 g^2_x +
12 \left(
\tilde V_f P^2 - \tilde U_{af} q^2 \right)^2 \right],\nonumber
\eea

\bea
&&
\tilde V_f=\frac{V_f}{V}\,, \quad \tilde V_{af} = \frac{V_{af}}{V}\,, \qquad
\ \tilde U_f=\frac{U_f}{V}\,, \quad \tilde U_{af} = \frac{U_{af}}{V}\,, \non
&&
V_f = \frac{1}{2} \sum_{j=1}^N \lp V_{ij}^{11}+V_{ij}^{12}\rp, \qquad
V_{af} = \frac{1}{2} \sum_{j=1}^N \lp V_{ij}^{11}-V_{ij}^{12}\rp\,, \non
&&
U_f = \frac{1}{2} \sum_{j=1}^N \lp U_{ij}^{11}+U_{ij}^{12}\rp, \qquad
U_{af} = \frac{1}{2} \sum_{j=1}^N \lp U_{ij}^{11}-U_{ij}^{12}\rp\,, \non
&&
\hspace{1.7mm} V = V_{ii}^{12}\,, \quad U = U_{ii}^{12}\,.
\eea
\vspace{-5mm}
\bea
&&J_1 = \sum_{t=1}^\infty (-1)^t 2 \ln \left(
1-n+n\re^{-2\beta V t}\right),\non
&&J_2 = \sum_{t=1}^\infty (-1)^t 4t
\ln \left( 1-n+n\re^{-2\beta V t}\right),
\eea
$\beta=(kT)^{-1}$, $k$ is the Boltzmann constant, $T$ is the absolute temperature.

The form of $J_1(n,T)$ and $J_2(n,T)$ functions is determined by the binomial random distribution of different types of ``particles''
(Rb or NH$_4$) and the replica method of configuration averaging. $J_1(n,T)$ is proportional to the concentration $n$, but
$J_2(n,T)$ reflects the correlation between both subsystems and naturally tends to zero in the limits $n\rightarrow 0$
and $n\rightarrow 1$.

Ferroelectric $P$, antiferroelectric $q$ order parameters and $g_z$, $g_x$ nearest neighbours
pair correlation functions are defined in the following way:
\be
\label{eq:2.7}
P =
\overline{<\sigma_{i1}>}, \qquad q = \overline{<\xi_{i1}>}, \qquad g_z
= \overline{<\sigma_{i1}\sigma_{i2}>}, \qquad g_x =
\overline{<\xi_{i1}\xi_{i2}>}, \ee
where $<\ldots>$ denotes thermal averaging for a given distribution of $z$- and $x$-oriented dipole momenta
of Rb-AsO$_4$ and NH$_4$-As$O_4$ groups, respectively, whereas a bar denotes a stochastic averaging over
different random distributions.

From the principle of stationarity for the free energy (\ref{eq:2.3}) under order parameters $P$, $q$ and both pair correlation functions
$g_z$ and $g_x$ (they determine a structure of nearest neighbours particles) the set of equations is obtained:
\bea
\label{eq:2.8}
P &=& \frac{\sinh(2B_1+\beta E\cos\Theta)}{\cosh(2B_1+\beta E \cos \Theta)+\re^{-2A_1}}\,,\non
q &=& \frac{\sinh (2B_2)}{\cosh (2B_2) + \re^{2A_2} \cosh(\beta E \sin\Theta)}\,,\non
g_z &=& \frac{-1+\re^{2A_1}\cosh(2B_1+\beta E \cos\Theta)}{1+\re^{2A_1}\cosh(2B_1+\beta E\cos\Theta)}\,,\non
g_x &=& \frac{\cosh(\beta E\sin\Theta)-\re^{-2A_2}\cosh (2B_2)}{\cosh(\beta E \sin \Theta)+\re^{-2A_2}\cosh (2B_2)}\,.
\eea

Spontaneous values (in zero external field $E$) of $P, q, g_z, g_x$ parameters are limited by $\pm 1$.
So, naturally, they may be interpreted as a set of order parameters of the system under investigation.

\section{Phase diagram. Dipole glass parameter}

\begin{figure}[htbp]
\centerline{
\includegraphics[width=0.65\textwidth]{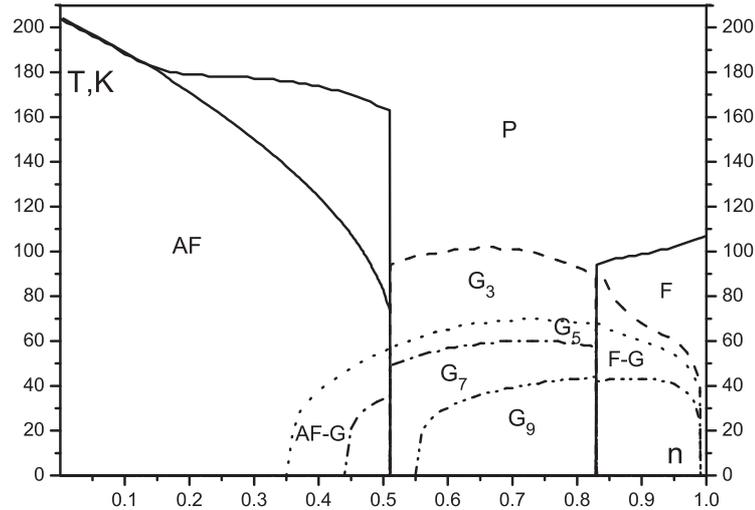}
}
\caption{
Extended phase diagram of RADA type compound built for the model Hamiltonian parameters:
$V_f = 70$~K, $U_f = -40$~K, $V_{af} = -80$~K, $U_{af} = 120$~K, $V = 65$~K, $U = -210$~K.
}
\label{fig1}
\end{figure}

Computing the equations (\ref{eq:2.8})  one can find the dependencies of $P$, $q$, $g_z$, $g_x$ parameters on temperature $T$ and concentration $n$.
Those dependencies form a phase diagram of the investigated ferro-antiferro\-electric mixed system in the plane $n-T$.
The principal outline of a phase diagram for RADA system was obtained in the paper~\cite{28}, but due to its considerable role
in the analysis of the mixed system behaviour, we reproduced it here (figure~\ref{fig1}).

The main features of the phase diagram are: 1) the existence of two non-crossing areas for ferroelectric (F) and antiferroelectric (AF)
phases; 2) the existence of pure dipole glass phase at intermediate concentration and relatively small temperature (this phase is divided
into several (G$_3$-G$_9$) subphases according to the order of the chaotic state); 3) the possibility of mixed ferroelectric-dipole glass (F-G)
and antiferroelectric-dipole glass (AF-G) phases existence. It should be noted that the upper dipole glass phase G$_3$ is rather a precursor
of the real dipole glass phase appearance. This conclusion is in good agreement with the experiment~\cite{32,33}.

The difference between RADA phase diagram in~\cite{28} and the one presented here is the line dividing the paraelectric phase and the antiferroelectric phase (P-AF).
In figure~\ref{fig1} the upper line represents the points of stable non-zero root arising for antiferroelectric order parameter $q$ while in~\cite{28}
the corresponding line indicates the points of its bifurcation (lower line in figure~\ref{fig1}). It should be noted that the only two fitting points were used to build
figure~\ref{fig1}. The first one is the antiferroelectric phase transition point temperature for pure NH$_4$H$_2$AsO$_4$ crystal, and the second one is
the ferroelectric phase transition point temperature for pure RbH$_2$AsO$_4$ crystal.

The dipole glass phase is interpreted as a possibility of different arrangements for the particles of nearest neighbours.
This fact is reflected by numerous solutions for pair correlation functions $g_z$ and $g_x$ at different values of temperature and concentration.
When the temperature decreases, the dipole glass state becomes more complicated.

In this context, the behaviour of correlation functions $g_z$, $g_x$ is characteristic of the dipole glass structure. However, there are a some
peculiarities connected with their relation to dipole glass state parameter. Usually~\cite{34}, this parameter is introduced to ferroelectrics
theory by analogy with spin glass parameter in ferromagnetic systems~\cite{16}, named Edwards and Anderson parameter. Edward and Anderson parameter is defined as a
second moment of magnetic order parameter. This moment is maximum for low temperature and vanishes at relatively large temperature $T_\mathrm{g}$ named
glassy temperature. As for ferroelectric systems, the situation is not that simple. The recent investigations~\cite{25} of mixed hydrogen bonded
ferro-antiferroelectric systems have shown some restriction to using the Edwards and Anderson parameter
for a complete definition of a dipole glass state. The second moment of ferroelectric order parameter does not vanish even for very high temperatures when no
glass state can exist.

Pair correlation functions $g_z$ and $g_x$, which are analogous to the second moment of order parameter, are nonzero even in paraelectric
phase $P$ (see figure~\ref{fig1}) and tend to zero for very high temperatures (theoretically $\lim_{T\rightarrow\infty} g_z = \lim_{T\rightarrow\infty} g_x = 0$).
According to our concept, the dipole glass phase is a state with numerous types of short-range correlations between particles, or numerous numbers of
solutions for a set of equations~(\ref{eq:2.8}). The intensity of correlations (values of $g_z$ and $g_x$) is important but not decisive for
dipole glass phase characterization. Dipole glass phase arises only at the moment when the number of different solutions for $g_z$ or $g_x$
changes from 1 to a bigger number. Such a situation is observed on the upper line of G$_3$ area (figure~\ref{fig1}).
As we have calculated (the phase diagram has been built on this basis), there
is only one solution for ($g_z$, $g_x$) pair in the paraelectric phase ($P$), three solutions for ($g_z$, $g_x$) pair in the G$_3$ area,
five solutions for ($g_z$, $g_x$) pair in the G$_5$ area and so on. With an increase of the number
of different solutions for $g_z$, $g_x$,  a bigger
number of different types of correlations of the nearest particles are observed. As a result, a more complicated arrangement of dipoles all over the crystal takes place.
This situation is accompanied by a more chaotic distribution of correlations and by the formation of a dipole glass phase.
When temperature decreases, a higher stage of dipole glass phase is realized. Thus, dipole glass parameter should take into account two factors:
the number of solutions for $g_z$, $g_x$ parameters and their absolute values.
It is obvious that the unique (trivial) solution for $g_z$, $g_x$,
which remains in the paraelectric phase must be eliminated.

Therefore, since both $g_z$ and $g_x$ functions form a dipole glass parameter $G$, we propose the next expression for its calculation:
\be
G = \frac{1}{2M_0} \sum_{i=1}^{m-1} \lb n|g_z^{(i)}| + (1-n)|g_x^{(i)}|\rb,
\ee
where $g_z^{(i)}$, $g_x^{(i)}$ are the roots of a set of equations (\ref{eq:2.8}), $m$ is the number of these roots for any temperature and concentration,
$N_0$ is the maximum number of roots, which are taken into account (in the presented calculations $M_0=9$), $n$ is the concentration.

The ``partial'' dipole glass parameters $G_z$ and $G_x$:
\bea
&&
G_z = \frac{1}{M_0} \sum_{i=1}^{m-1} |g_z^{(i)}|\,,\non
&&
G_x = \frac{1}{M_0} \sum_{i=1}^{m-1} |g_x^{(i)}|
\eea
describe the power of dipole glass state intensity according to $z-z$ and $x-x$ correlation of Rb-AsO$_4$ and NH$_4$-AsO$_4$ dipoles, correspondingly.

\begin{figure}[!t]
\centerline{
\includegraphics[width=0.65\textwidth]{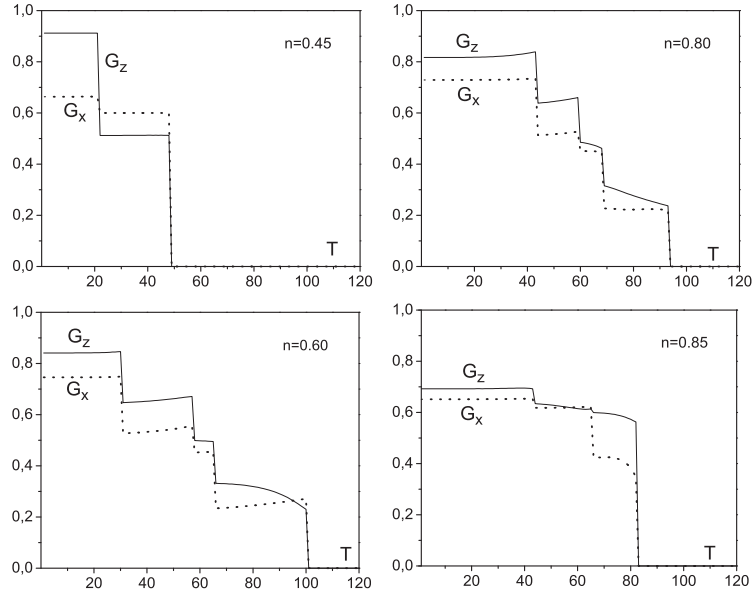}
}
\caption{
Temperature dependencies of the ``partial'' dipole glass parameters $G_z$ and $G_x$ for different concentrations.
}
\label{fig2}
\end{figure}

\begin{figure}[!b]
\centerline{
\includegraphics[width=0.65\textwidth]{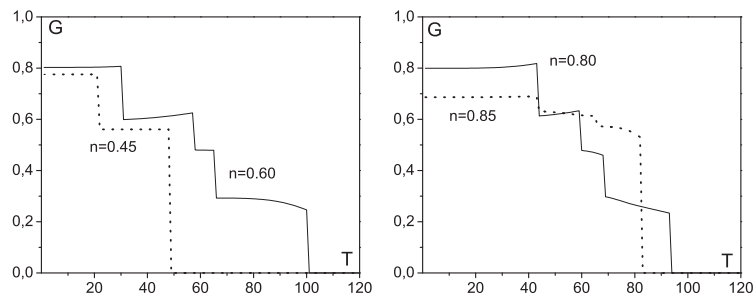}
}
\caption{
Temperature dependencies of the dipole glass parameters $G$ for different concentrations.
}
\label{fig3}
\end{figure}

The temperature dependencies for $G_z$ and $G_x$ at different concentrations are presented in figure~\ref{fig2}, and for $G$ they are presented in figure~\ref{fig3}.
The most characteristic feature in the behaviour of $G_z$, $G_x$ and $G$ are their step-like dependence. The multigraded character of a dipole glass parameter is caused
by the complex nature of dipole glass phase to exactly change the number of roots for $g_z$, $g_x$ correlation functions (see figure~\ref{fig1}).
The position of ``jumps'' for $G_z$, $G_x$, $G$ is determined by the lines dividing the areas of P-G$_3$, G$_3$-G$_5$, \ldots{} for ``pure'' dipole glass phases and F-G$_3$, G$_3$-G$_4$, \ldots;
AF-G$_3$, G$_3$-G$_5$, \ldots{} for the mixed ordered and dipole glass phases. The existence of those lines is determined by the finite number of different roots of equations~(\ref{eq:2.8})
in every area of temperature and concentration.
Theoretically, this number is infinite~\cite{32,35}, so the step-like behaviour of dipole glass parameter in these limits may transform into a quasi-continuous curve.

\section{Conclusions}

A new approach to the calculation of the dipole glass parameter in mixed ferro-antiferroelectric systems  has been developed.
It is based on the conception of a decisive role of the pair correlations of the nearest neighbours particles in a crystalline
lattice for the dipole glass phase formation. The corresponding correlation function has been calculated in the second order of replica expansion
and using the cluster approximation. This microscopic approach leads to the basic physical statement: the dipole glass phase is a polystate arrangement of the nearest
neighbours particles with the possibility of non-equilibrium occupation of them. In such  a way, the proposed definition differs from the well known Edwards and Anderson spin glass parameter in the
quenched ferromagnetic system theory.

The calculated dipole glass parameter takes into account not only the intensity of interparticle correlations but also the multiplicity of a
different type of those correlations. The
latter corresponds to different possible types of short-range order in the mixed system and usually describes a metastable states.
The total dipole glass parameter is formed both by ferro-ferro and antiferro-antiferro types of correlations. The step-like behaviour of
the dipole glass parameter reflects the non-uniform
structure of the dipole glass phase and the dependence of the intensity of the disorder on temperature and concentration.
When the number of different roots for pair correlation functions increases, the temperature dependence of this parameter becomes more smooth.

The obtained phase diagram ``temperature-concentration'' demonstrates the possibility of the existence of both pure ferroelectric, antiferroelectric,
dipole glass phases and mixed ferroelectric-dipole glass, antiferroelectric-dipole glass phases. This fact was recently confirmed
by numerous experimental investigations. The precursors of the dipole glass phase which appears at rather high temperatures
are interpreted as the state with the finite number of different types of interparticle correlations, in counterbalance to paraelectric state with only one type of
those correlations.

\ukrainianpart

\title{Поведінка параметра дипольного скла для сегнето-антисегнетоелектричних твердих сумішей}

\author{М.А.Кориневський\refaddr{label1,label2,label3}, В.Б.Солов'ян\refaddr{label1}}

\addresses{
\addr{label1}Інститут фізики конденсованих систем НАН України,
          вул. І.~Свєнцiцького, 1, 79011 Львiв, Україна
\addr{label2} Національний університет ``Львівська Політехніка'', вул. С.~Бандери 12, 79013 Львів, Україна
\addr{label3} Інститут фізики, Щецінський університет, вул. Вєлькопольска 15, 70451 Щецін, Польща
}

\makeukrtitle

\begin{abstract}
\tolerance=3000%
Запропоновано нове означення параметра дипольного скла для сегнето-антисегнетоелектричних твердих сумішей.
Цей параметр будується на парних кореляційних функціях взаємодіючих дипольних моментів
частинок найближчих сусідів. Розраховано та досліджено поведінку параметра дипольного скла, а також фізичні властивості фази дипольного скла.
\keywords сегнетоелектрики, антисегнетоелектрики, змішані системи, кореляційні функції, дипольне скло

\end{abstract}

\end{document}